\documentclass[aps,prl,preprint,groupedaddress,showkeys,showpacs]{revtex4-1}

\usepackage{graphicx}
\usepackage{dcolumn}
\usepackage{bm}

\begin{document}

\title{Bistability loss as key feature in azobenzene (non-)switching on metal surfaces}

\author{Reinhard J. Maurer}
 \email{reinhard.maurer@ch.tum.de}
\author{Karsten Reuter}

\affiliation{
Department Chemie, Technische Universit\"at M\"unchen,\\
Lichtenbergstrasse 4, 85747 Garching b. M\"unchen, Germany
}

\date{\today}

\begin{abstract}
Coinage metal adsorbed azobenzene is investigated as prototypical molecular switch. It is shown that switching capabilities are not just lost due to excited state quenching, but already due to changes in the ground state energetics. Electron demanding coadsorbates are suggested as strategy to regain the switching function.
\end{abstract}


\maketitle

Stimulating a controlled conformational change of functional molecules adsorbed on solid surfaces is a key goal of molecular nanotechnology. Along this route molecular switches, i.e. molecules that can be reversibly switched between metastable states in gas-phase or solution, have been a primary target. \cite{Morgenstern2011}
Unfortunately, even at most unreactive close-packed noble metal surfaces such molecules generally do not show capabilities for light or electron induced switching. Molecular functionalization aiming to further decouple the photochromic moiety, both spatially and electronically, has been a standard strategy to regain the switching function, viewing excited state quenching at the metal surface as central reason for the loss of function upon adsorption. A prominent example is tetra-tert-butyl-azobenzene (TBA), which in contrast to its parent molecule azobenzene (Ab) can be successfully switched on Au(111). \cite{Comstock2008, Hagen2007}

Notwithstanding, the hitherto unsuccessful attempts to switch both Ab and TBA at the closely related Ag(111) surface indicate that factors other than photochromic decoupling might also be important. Several explanations why switching of TBA on other metal surfaces is not observed have been formulated \cite{Wolf2009,Bronner2012}, but no unified picture specifying how surface interaction changes the stability and reactivity of metal-mounted azocompounds has yet been reached. Recalling that the existence of two metastable states is a fundamental prerequisite for switching, it is intriguing to realize that no study has hitherto addressed the possibility that the absence of surface mounted switching could as well simply stem from a changed ground-state stability.

On the basis of the well understood Ab gas-phase isomerization mechanisms, we set out to investigate this point by studying the effect of coinage metal surface adsorption on the ground-state barriers with dispersion-corrected density-functional theory (DFT) calculations. \cite{Clark2005, Perdew1996, McNellis2009, Tkatchenko2009} Our results suggest that surface adsorption modifies the azobenzene groundstate stability in a way to de facto remove bistability and therewith allow immediate thermal re-isomerization from the previously metastable state. Quite naturally this calls for a paradigm change in molecular switch design strategies. In addition to a controlled decoupling of the photochromic moiety, functionalization needs to target a balanced stabilization of all structures involved in the isomerization in order to preserve the aspired switching function upon surface mounting.

\begin{figure}
\includegraphics{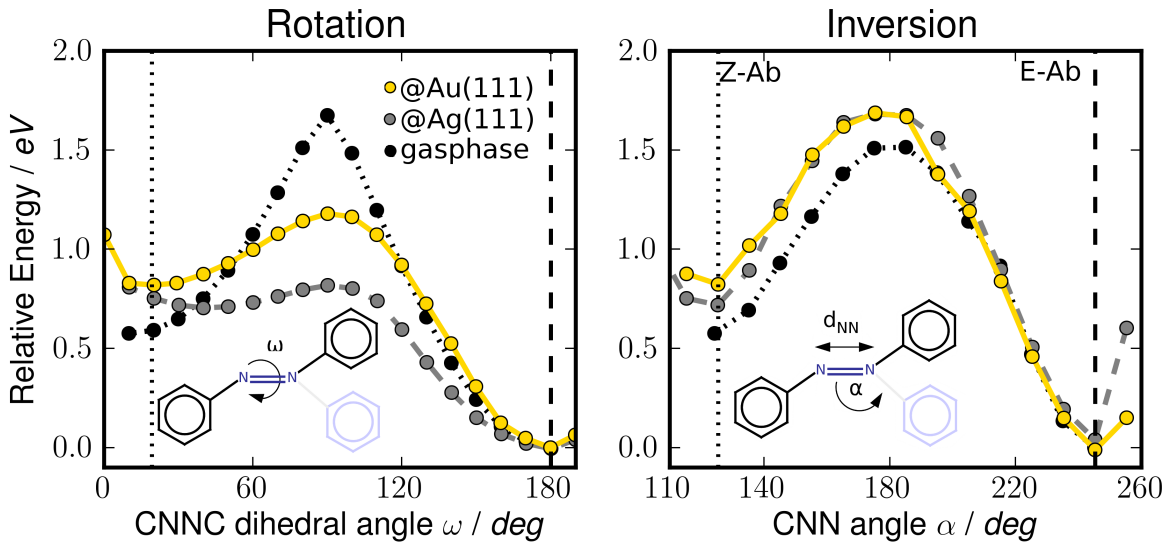}
\caption{\label{fig1} Minimum energy isomerization paths along the rotation (left) and inversion (right) coordinate. Shown are ground state curves for the molecule in gasphase (dotted line), adsorbed on a Ag(111) surface (dashed line) and adsorbed on a Au(111) surface (solid line).}
\end{figure}

The vast amount of mechanistic studies of Ab isomerization in gas-phase and solution mainly focuses on two different mechanisms \cite{Conti2008, Crecca2006, Cembran2004, Cattaneo1999,Maurer2011}: A dihedral rotation of one phenyl group around the central azo-bridge and an initially planar inversion around one of the CNN bond angles, cf. Fig. \ref{fig1}. Present consensus points towards a dominance of the rotational mechanism upon photoexcitation. \cite{Shao2008,Bockmann2010} 
Nonetheless, in the ground state both mechanisms show significant barriers, 1.8 and 1.5\,eV for rotation and inversion, respectively, as seen from the more stable E-Ab state \cite{Maurer2011}. The rotational barrier maximum also coincides with a state crossing with the first singlet excited state, explaining the conical peak feature well reproduced by our DFT calculations\cite{Cembran2004,Maurer2011}. Recomputing these barriers for Ab adsorbed on Ag(111) we find the inversion barrier almost unchanged. In contrast and as shown in Fig. \ref{fig1}, the rotational barrier is drastically reduced by about 1\,eV with respect to the E-Ab isomer and the conical peak shape is gone. In addition, the metastable Z-Ab isomer changes its minimum energy geometry towards higher dihedral angle and destabilizes by another 0.2\,eV as compared to E-Ab, i.e. while Z-Ab is 0.5\,eV higher in energy in the gas-phase, it is 0.7\,eV at the surface. These effects together leave the re-isomerization from Z-Ab back to the more stable E-Ab with a minimal zero-point energy corrected barrier of 50\,meV, while simultaneously minimizing potential restrictions in the DFT description of the barrier region. Without even entering into the details of the light or electron driven excitation per se, already this insignificant barrier alone is enough to explain why Ab at Ag(111) will not switch - the fundamental bistability prerequisite to the switching function is simply lost. In particular, even if a very efficient photo-excitation mechanism to the cis state existed, the strongly vibrationally activated ground-state Z-Ab isomer resulting from the isomerization would still not be sufficiently stable to be observed.

\begin{figure}
\includegraphics[width=8cm]{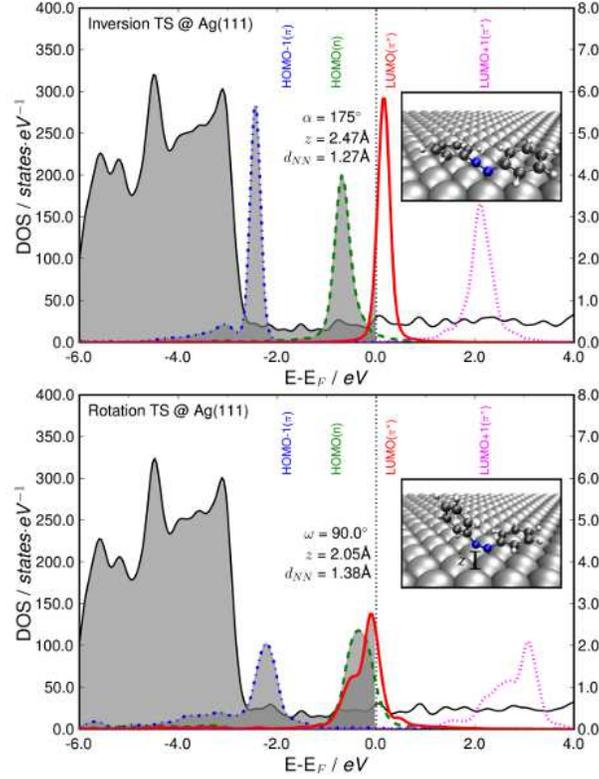}
\caption{\label{fig2} Density-of-states (DOS) for the TS along inversion (top) and rotation (bottom) isomerization of Ab on Ag(111) surface. Also shown is the DOS projected onto the molecular frontier orbitals, as well as the TS geometries including important structural parameters.}
\end{figure}

\begin{figure}
\includegraphics{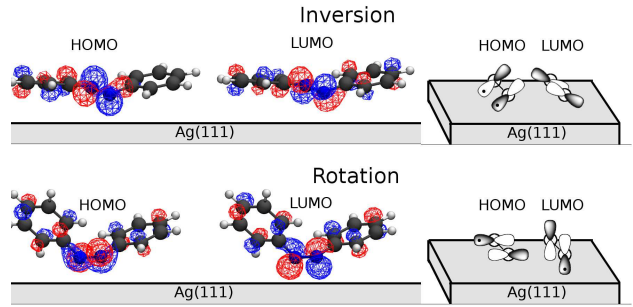}
\caption{\label{fig3} HOMO and LUMO orbital shape of Ab at Ag(111) for the inversion (top) and rotational (bottom) TS. Shown are both calculated isosurfaces (left) and schematic representations (right).}
\end{figure}

The reason for the strong preferential reduction of the rotational barrier lies in the formation of a strong chemisorption bond along this pathway. Whereas at both minimum energy structures and the inversion transition state (TS) the surface stabilization results to more than 90 \% just from dispersive interactions, the adsorption energy at the rotational TS comes to 40 \% from the semi-local DFT functional. Geometrically this covalent bond contribution at the TS is indicated by a significant reduction of the vertical height and an elongation of the bond length of the central azo-bridge, from 2.95 \AA\ at E-Ab to 2.05 \AA\ and from 1.31 \AA\ at E-Ab to 1.38 \AA\ , respectively. In the orbital-projected density-of-states (DOS) shown in Fig. \ref{fig2}, this different surface interaction at inversion and rotational TS can also nicely be discerned. At the inversion TS, the frontier orbitals exhibit only the same minimal broadening due to the interaction with the metal bands as had been found before for the two minimum energy structures \cite{McNellis2009a}. The small amount of charge transferred to the lowest-unoccupied molecular orbital (LUMO) shows that in all these cases the surface interaction can be understood in the classic Dewar-Chatt-Duncanson $\pi$-donor-$\pi$*-acceptor model for bonding between metals and conjugated organics\cite{Dewar1951,Chatt1953}. Along the rotational path the frontier orbitals show instead a much stronger broadening and splitting. At the TS the LUMO and highest-occupied molecular orbital (HOMO) are in fact almost degenerate and situated slightly below the Fermi level. The situation is thus highly reminiscent of a diradical state, known to be a highly active chemical intermediate for example in Diels-Alder reactions\cite{Salem1972}. Visualizing the molecular frontier orbitals at the surface adsorbed geometry, as done in Fig. \ref{fig3},then immediately shows that both, HOMO and LUMO, orbitals have a nonbonding character and are perfectly arranged to interact with localized metal $d$ states, as opposed to the orbitals found in the inversion barrier structure.

To paraphrase this, following rotation from the E-Ab isomer the azo-bridge double bond is increasingly weakened and the two orbitals develop a non-bonding, diradical character. This reactive state is stabilized by efficiently accepting electron charge from the underlying metal. This stabilization is much stronger than the overall stabilization due to dispersive interaction with the substrate. The latter increases only slightly from Z-Ab to E-Ab due to the increasingly parallel alignment of the two phenyl rings. Altogether this unbalanced way in which particularly Ab in the rotational TS geometry interacts with the surface essentially eliminates the bistability feature of the molecule and therewith impairs the switching function.

One way to regain a more balanced adsorption of all geometries is by reducing the electron availability and therewith the stabilization of the diradical state through charge transfer. On the substrate side this is effectively achieved by a lowered Fermi energy level (higher work function), as e.g. realized at the Au(111) surface. Indeed, this partly re-establishes the rotational barrier as shown in Fig. \ref{fig1}. The resulting depth of 0.38\,eV for the metastable Z-Ab basin is still much reduced compared to the gas-phase (1.02\,eV) though. The experimentally observed electron-induced switching using an STM tip does in fact reveal the existence of a barrier on this surface\cite{Comstock2005}. In the case of photo-induced switching, vibronic coupling is very strong and energy of the order of eVs can be efficiently transferred from electronic to vibrational degrees of freedom. A reduced barrier of 0.38\,eV might thus not be completely prohibitive to thermal back-reaction of a vibrationally hot cis isomer, and would then be one possible rationalization for the lacking photo-switching function of Ab at this surface. Another route to achieve a more balanced binding of all involved geometries is via molecular functionalization. However, this must follow different design strategies than those merely focused on a decoupling of the photochromic moiety, e.g. in the present case the symmetric addition of tert-butyl groups in TBA. In fact, we calculate a Z-TBA basin at Ag(111) about as shallow as that of Z-Ab at Ag(111) and Au(111), i.e. that the rotational TS bonding is barely affected by the functionalization.

In conclusion we generalize that thermal isomerization of surface-adsorbed molecules including double bond twisting events proceeds via barrier geometries that couple much stronger to the surface than the minimum energy or other transition state structures. The ensuing lowering of the ground-state barriers might be sufficiently strong to eliminate the bistability prerequisite for switching, as illustrated here for azobenzene at Ag(111). Without doubt, molecular functionalization must centrally target a tuned interaction of the photochromic moiety with the underlying metal to prevent ultra-fast quenching of excited states important for the isomerization. However, as shown here, a second target must also be to achieve a balanced surface interaction of all geometries involved in the isomerization process. For the present case of azobenzene-derivatives at noble metal surfaces this means to specifically aim at a selective destabilization of the diradical rotational transition state. As an intriguing and hitherto not pursued route this could be achieved by further reducing the substrate electron availability e.g. through electron demanding coadsorbates that increase the work function.

\subsection{Methods Section}

For our calculations we employ the pseudopotential plane wave code CASTEP 5.5.1 \cite{Clark2005} using standard library ultrasoft pseudopotentials\cite{Vanderbilt1990}. Electronic exchange and correlation were treated with the  semi-local PBE functional \cite{Perdew1996} and van der Waals interactions are accounted for via a semi-empirical dispersion correction scheme\cite{McNellis2009,Tkatchenko2009}. The detailed system setup, computational parameters and convergence behaviour were described already in a previous publication\cite{McNellis2009a}. In short, all calculations were performed with (6x4) and (6x5) frozen (111) 4 layer surface slabs of Ag and Au with 350 eV or 450 eV plane wave cutoff for Azobenzene and 3,3'5,5'-tetra-tert-butyl-azobenyene (TBA), respectively. The vacuum was chosen to exceed 20\,\AA. All energy differences and adsorption energies were calculated at a 8x4x1 Monkhorst-Pack grid\cite{Monkhorst76} for Azobenzene systems and a 6x4x1 Monkhorst-Pack grid for TBA systems. Relative energies and adsorption energies are converged to $\pm$20\,meV. We note that the employed supercell approach has been rigorously evaluated in previous work \cite{McNellis2009a,Mercurio2010,McNellis2010} and is known to accurately reproduce geometries, but overestimate absolute adsorption energies due to the neglect of electronic screening in the dispersion correction \cite{McNellis2010}. For this reason we have recalculated the minimum and transition state energetics also with the screened dispersion correction approach recently introduced by Ruiz et al. \cite{Ruiz2012}, without obtaining qualitative changes to the results reported in this work. We expect a similarly robust description of the covalent azo-metal interaction, considering the well-established accuracy of the employed PBE functional for nitrogen-based catalytic reactions at coinage metals.

\begin{acknowledgments}
The authors gratefully acknowledge technical support by Christoph Scheurer and J\"org Meyer, as well as computing time at the Leibniz Supercomputing Center (LRZ) and the Garching Supercomputing Center (RZG) of the Max-Planck-Society (MPG). This work was funded by the German Research Foundation (DFG). 
\end{acknowledgments}

%


\begin{thebibliography}{10}%
\makeatletter
\providecommand \@ifxundefined [1]{%
 \ifx #1\undefined \expandafter \@firstoftwo
 \else \expandafter \@secondoftwo
\fi
}%
\providecommand \@ifnum [1]{%
 \ifnum #1\expandafter \@firstoftwo
 \else \expandafter \@secondoftwo
\fi
}%
\providecommand \enquote [1]{``#1''}%
\providecommand \bibnamefont  [1]{#1}%
\providecommand \bibfnamefont [1]{#1}%
\providecommand \citenamefont [1]{#1}%
\providecommand\href[0]{\@sanitize\@href}%
\providecommand\@href[1]{\endgroup\@@startlink{#1}\endgroup\@@href}%
\providecommand\@@href[1]{#1\@@endlink}%
\providecommand \@sanitize [0]{\begingroup\catcode`\&12\catcode`\#12\relax}%
\@ifxundefined \pdfoutput {\@firstoftwo}{%
 \@ifnum{\z@=\pdfoutput}{\@firstoftwo}{\@secondoftwo}%
}{%
 \providecommand\@@startlink[1]{\leavevmode}%
 \providecommand\@@endlink[0]{}%
}{%
 \providecommand\@@startlink[1]{%
  \leavevmode
  \pdfstartlink
   attr{/Border[0 0 1 ]/H/I/C[0 1 1]}%
   user{/Subtype/Link/A<</Type/Action/S/URI/URI(#1)>>}%
  \relax
 }%
 \providecommand\@@endlink[0]{\pdfendlink}%
}%
\providecommand \url  [0]{\begingroup\@sanitize \@url }%
\providecommand \@url [1]{\endgroup\@href {#1}{\urlprefix}}%
\providecommand \urlprefix [0]{URL }%
\providecommand \Eprint[0]{\href }%
\@ifxundefined \urlstyle {%
  \providecommand \doi [1]{doi:\discretionary{}{}{}#1}%
}{%
  \providecommand \doi [0]{doi:\discretionary{}{}{}\begingroup
  \urlstyle{rm}\Url }%
}%
\providecommand \doibase [0]{http://dx.doi.org/}%
\providecommand \Doi[1]{\href{\doibase#1}}%
\providecommand \bibAnnote [3]{%
  \BibitemShut{#1}%
  \begin{quotation}\noindent
    \textsc{Key:}\ #2\\\textsc{Annotation:}\ #3%
  \end{quotation}%
}%
\providecommand \bibAnnoteFile [2]{%
  \IfFileExists{#2}{\bibAnnote {#1} {#2} {\input{#2}}}{}%
}%
\providecommand \typeout [0]{\immediate \write \m@ne }%
\providecommand \selectlanguage [0]{\@gobble}%
\providecommand \bibinfo [0]{\@secondoftwo}%
\providecommand \bibfield [0]{\@secondoftwo}%
\providecommand \translation [1]{[#1]}%
\providecommand \BibitemOpen[0]{}%
\providecommand \bibitemStop [0]{}%
\providecommand \bibitemNoStop [0]{.\EOS\space}%
\providecommand \EOS [0]{\spacefactor3000\relax}%
\providecommand \BibitemShut [1]{\csname bibitem#1\endcsname}%
\bibitem{Morgenstern2011}%
  \BibitemOpen
  \bibfield{author}{%
  \bibinfo {author} {\bibfnamefont{K.}~\bibnamefont{Morgenstern}},\ }%
  \bibfield{journal}{%
  \bibinfo {journal} {Progr. Surf. Sci.}\ }%
  \textbf{\bibinfo {volume} {86}},\ \bibinfo {pages} {115} (\bibinfo {year}
  {2011})%
  \bibAnnoteFile{NoStop}{Morgenstern2011}%
\bibitem{Comstock2008}%
  \BibitemOpen
  \bibfield{author}{%
  \bibinfo {author} {\bibfnamefont{M.~J.}\ \bibnamefont{Comstock}}, \bibinfo
  {author} {\bibfnamefont{N.}~\bibnamefont{Levy}}, \bibinfo {author}
  {\bibfnamefont{J.}~\bibnamefont{Cho}}, \bibinfo {author}
  {\bibfnamefont{L.}~\bibnamefont{Berbil-Bautista}}, \bibinfo {author}
  {\bibfnamefont{M.~F.}\ \bibnamefont{Crommie}}, \bibinfo {author}
  {\bibfnamefont{D.~A.}\ \bibnamefont{Poulsen}},\ and\ \bibinfo {author}
  {\bibfnamefont{J.~M.~J.}\ \bibnamefont{Fréchet}},\ }%
  \bibfield{journal}{%
  \bibinfo {journal} {Appl. Phys. Lett.}\ }%
  \textbf{\bibinfo {volume} {92}},\ \bibinfo {pages} {123107} (\bibinfo {year}
  {2008})%
  \bibAnnoteFile{NoStop}{Comstock2008}%
\bibitem{Hagen2007}%
  \BibitemOpen
  \bibfield{author}{%
  \bibinfo {author} {\bibfnamefont{S.}~\bibnamefont{Hagen}}, \bibinfo {author}
  {\bibfnamefont{F.}~\bibnamefont{Leyssner}}, \bibinfo {author}
  {\bibfnamefont{D.}~\bibnamefont{Nandi}}, \bibinfo {author}
  {\bibfnamefont{M.}~\bibnamefont{Wolf}},\ and\ \bibinfo {author}
  {\bibfnamefont{P.}~\bibnamefont{Tegeder}},\ }%
  \bibfield{journal}{%
  \bibinfo {journal} {Chem. Phys. Lett.}\ }%
  \textbf{\bibinfo {volume} {444}},\ \bibinfo {pages} {85} (\bibinfo {year}
  {2007})%
  \bibAnnoteFile{NoStop}{Hagen2007}%
\bibitem{Wolf2009}%
  \BibitemOpen
  \bibfield{author}{%
  \bibinfo {author} {\bibfnamefont{M.}~\bibnamefont{Wolf}}\ and\ \bibinfo
  {author} {\bibfnamefont{P.}~\bibnamefont{Tegeder}},\ }%
  \bibfield{journal}{%
  \bibinfo {journal} {Surf. Sci.}\ }%
  \textbf{\bibinfo {volume} {603}},\ \bibinfo {pages} {1506} (\bibinfo {year}
  {2009})%
  \bibAnnoteFile{NoStop}{Wolf2009}%
\bibitem{Bronner2012}%
  \BibitemOpen
  \bibfield{author}{%
  \bibinfo {author} {\bibfnamefont{C.}~\bibnamefont{Bronner}}, \bibinfo
  {author} {\bibfnamefont{M.}~\bibnamefont{Schulze}}, \bibinfo {author}
  {\bibfnamefont{S.}~\bibnamefont{Hagen}},\ and\ \bibinfo {author}
  {\bibfnamefont{P.}~\bibnamefont{Tegeder}},\ }%
  \bibfield{journal}{%
  \bibinfo {journal} {New J. Phys.}\ }%
  \textbf{\bibinfo {volume} {14}},\ \bibinfo {pages} {043023} (\bibinfo {year}
  {2012})%
  \bibAnnoteFile{NoStop}{Bronner2012}%
\bibitem{Clark2005}%
  \BibitemOpen
  \bibfield{author}{%
  \bibinfo {author} {\bibfnamefont{S.}~\bibnamefont{Clark}}, \bibinfo {author}
  {\bibfnamefont{M.}~\bibnamefont{Segall}}, \bibinfo {author}
  {\bibfnamefont{C.}~\bibnamefont{Pickard}}, \bibinfo {author}
  {\bibfnamefont{P.}~\bibnamefont{Hasnip}}, \bibinfo {author}
  {\bibfnamefont{M.}~\bibnamefont{Probert}}, \bibinfo {author}
  {\bibfnamefont{K.}~\bibnamefont{Refson}},\ and\ \bibinfo {author}
  {\bibfnamefont{M.}~\bibnamefont{Payne}},\ }%
  \bibfield{journal}{%
  \bibinfo {journal} {Z. Kristallogr.}\ }%
  \textbf{\bibinfo {volume} {220}},\ \bibinfo {pages} {567} (\bibinfo {year}
  {2005})%
  \bibAnnoteFile{NoStop}{Clark2005}%
\bibitem{Perdew1996}%
  \BibitemOpen
  \bibfield{author}{%
  \bibinfo {author} {\bibfnamefont{J.~P.}\ \bibnamefont{Perdew}}, \bibinfo
  {author} {\bibfnamefont{K.}~\bibnamefont{Burke}},\ and\ \bibinfo {author}
  {\bibfnamefont{M.}~\bibnamefont{Ernzerhof}},\ }%
  \bibfield{journal}{%
  \bibinfo {journal} {Phys. Rev. Lett.}\ }%
  \textbf{\bibinfo {volume} {77}},\ \bibinfo {pages} {3865} (\bibinfo {year}
  {1996})%
  \bibAnnoteFile{NoStop}{Perdew1996}%
\bibitem{McNellis2009}%
  \BibitemOpen
  \bibfield{author}{%
  \bibinfo {author} {\bibfnamefont{E.~R.}\ \bibnamefont{McNellis}}, \bibinfo
  {author} {\bibfnamefont{J.}~\bibnamefont{Meyer}},\ and\ \bibinfo {author}
  {\bibfnamefont{K.}~\bibnamefont{Reuter}},\ }%
  \bibfield{journal}{%
  \bibinfo {journal} {Phys. Rev. B}\ }%
  \textbf{\bibinfo {volume} {80}},\ \bibinfo {pages} {205414} (\bibinfo {year}
  {2009})%
  \bibAnnoteFile{NoStop}{McNellis2009}%
\bibitem{Tkatchenko2009}%
  \BibitemOpen
  \bibfield{author}{%
  \bibinfo {author} {\bibfnamefont{A.}~\bibnamefont{Tkatchenko}}\ and\ \bibinfo
  {author} {\bibfnamefont{M.}~\bibnamefont{Scheffler}},\ }%
  \bibfield{journal}{%
  \bibinfo {journal} {Phys. Rev. Lett.}\ }%
  \textbf{\bibinfo {volume} {102}},\ \bibinfo {pages} {073005} (\bibinfo {year}
  {2009})%
  \bibAnnoteFile{NoStop}{Tkatchenko2009}%
\bibitem{Conti2008}%
  \BibitemOpen
  \bibfield{author}{%
  \bibinfo {author} {\bibfnamefont{I.}~\bibnamefont{Conti}}, \bibinfo {author}
  {\bibfnamefont{M.}~\bibnamefont{Garavelli}},\ and\ \bibinfo {author}
  {\bibfnamefont{G.}~\bibnamefont{Orlandi}},\ }%
  \bibfield{journal}{%
  \bibinfo {journal} {J. Am. Chem. Soc.}\ }%
  \textbf{\bibinfo {volume} {130}},\ \bibinfo {pages} {5216} (\bibinfo {year}
  {2008})%
  \bibAnnoteFile{NoStop}{Conti2008}%
\bibitem{Crecca2006}%
  \BibitemOpen
  \bibfield{author}{%
  \bibinfo {author} {\bibfnamefont{C.~R.}\ \bibnamefont{Crecca}}\ and\ \bibinfo
  {author} {\bibfnamefont{A.~E.}\ \bibnamefont{Roitberg}},\ }%
  \bibfield{journal}{%
  \bibinfo {journal} {J. Phys. Chem. A}\ }%
  \textbf{\bibinfo {volume} {110}},\ \bibinfo {pages} {8188} (\bibinfo {year}
  {2006})%
  \bibAnnoteFile{NoStop}{Crecca2006}%
\bibitem{Cembran2004}%
  \BibitemOpen
  \bibfield{author}{%
  \bibinfo {author} {\bibfnamefont{A.}~\bibnamefont{Cembran}}, \bibinfo
  {author} {\bibfnamefont{F.}~\bibnamefont{Bernardi}}, \bibinfo {author}
  {\bibfnamefont{M.}~\bibnamefont{Garavelli}}, \bibinfo {author}
  {\bibfnamefont{L.}~\bibnamefont{Gagliardi}},\ and\ \bibinfo {author}
  {\bibfnamefont{G.}~\bibnamefont{Orlandi}},\ }%
  \bibfield{journal}{%
  \bibinfo {journal} {J. Am. Chem. Soc.}\ }%
  \textbf{\bibinfo {volume} {126}},\ \bibinfo {pages} {3234} (\bibinfo {year}
  {2004})%
  \bibAnnoteFile{NoStop}{Cembran2004}%
\bibitem{Cattaneo1999}%
  \BibitemOpen
  \bibfield{author}{%
  \bibinfo {author} {\bibfnamefont{P.}~\bibnamefont{Cattaneo}}\ and\ \bibinfo
  {author} {\bibfnamefont{M.}~\bibnamefont{Persico}},\ }%
  \bibfield{journal}{%
  \bibinfo {journal} {Phys. Chem. Chem. Phys.: PCCP}\ }%
  \textbf{\bibinfo {volume} {1}},\ \bibinfo {pages} {4739} (\bibinfo {year}
  {1999})%
  \bibAnnoteFile{NoStop}{Cattaneo1999}%
\bibitem{Maurer2011}%
  \BibitemOpen
  \bibfield{author}{%
  \bibinfo {author} {\bibfnamefont{R.~J.}\ \bibnamefont{Maurer}}\ and\ \bibinfo
  {author} {\bibfnamefont{K.}~\bibnamefont{Reuter}},\ }%
  \bibfield{journal}{%
  \bibinfo {journal} {J. Chem. Phys.}\ }%
  \textbf{\bibinfo {volume} {135}},\ \bibinfo {pages} {224303} (\bibinfo {year}
  {2011})%
  \bibAnnoteFile{NoStop}{Maurer2011}%
\bibitem{Shao2008}%
  \BibitemOpen
  \bibfield{author}{%
  \bibinfo {author} {\bibfnamefont{J.}~\bibnamefont{Shao}}, \bibinfo {author}
  {\bibfnamefont{Y.}~\bibnamefont{Lei}}, \bibinfo {author}
  {\bibfnamefont{Z.}~\bibnamefont{Wen}}, \bibinfo {author}
  {\bibfnamefont{Y.}~\bibnamefont{Dou}},\ and\ \bibinfo {author}
  {\bibfnamefont{Z.}~\bibnamefont{Wang}},\ }%
  \bibfield{journal}{%
  \bibinfo {journal} {J. Chem. Phys.}\ }%
  \textbf{\bibinfo {volume} {129}},\ \bibinfo {pages} {164111} (\bibinfo {year}
  {2008})%
  \bibAnnoteFile{NoStop}{Shao2008}%
\bibitem{Bockmann2010}%
  \BibitemOpen
  \bibfield{author}{%
  \bibinfo {author} {\bibfnamefont{M.}~\bibnamefont{B\"{o}ckmann}}, \bibinfo
  {author} {\bibfnamefont{N.~L.}\ \bibnamefont{Doltsinis}},\ and\ \bibinfo
  {author} {\bibfnamefont{D.}~\bibnamefont{Marx}},\ }%
  \bibfield{journal}{%
  \bibinfo {journal} {J. Phys. Chem. A}\ }%
  \textbf{\bibinfo {volume} {114}},\ \bibinfo {pages} {745} (\bibinfo {year}
  {2010})%
  \bibAnnoteFile{NoStop}{Bockmann2010}%
\bibitem{McNellis2009a}%
  \BibitemOpen
  \bibfield{author}{%
  \bibinfo {author} {\bibfnamefont{E.}~\bibnamefont{McNellis}}, \bibinfo
  {author} {\bibfnamefont{J.}~\bibnamefont{Meyer}}, \bibinfo {author}
  {\bibfnamefont{A.}~\bibnamefont{Baghi}},\ and\ \bibinfo {author}
  {\bibfnamefont{K.}~\bibnamefont{Reuter}},\ }%
  \bibfield{journal}{%
  \bibinfo {journal} {Phys. Rev. B}\ }%
  \textbf{\bibinfo {volume} {80}},\ \bibinfo {pages} {035414} (\bibinfo {year}
  {2009})%
  \bibAnnoteFile{NoStop}{McNellis2009a}%
\bibitem{Dewar1951}%
  \BibitemOpen
  \bibfield{author}{%
  \bibinfo {author} {\bibfnamefont{M.}~\bibnamefont{Dewar}},\ }%
  \bibfield{journal}{%
  \bibinfo {journal} {Bull. Soc. Chim. France}\ }%
  \textbf{\bibinfo {volume} {18}},\ \bibinfo {pages} {C79} (\bibinfo {year}
  {1951})%
  \bibAnnoteFile{NoStop}{Dewar1951}%
\bibitem{Chatt1953}%
  \BibitemOpen
  \bibfield{author}{%
  \bibinfo {author} {\bibfnamefont{J.}~\bibnamefont{Chatt}}\ and\ \bibinfo
  {author} {\bibfnamefont{L.~A.}\ \bibnamefont{Duncanson}},\ }%
  \bibfield{journal}{%
  \bibinfo {journal} {J. Chem. Soc.}\ }%
  \textbf{\bibinfo {volume} {1}},\ \bibinfo {pages} {2939} (\bibinfo {year}
  {1953})%
  \bibAnnoteFile{NoStop}{Chatt1953}%
\bibitem{Salem1972}%
  \BibitemOpen
  \bibfield{author}{%
  \bibinfo {author} {\bibfnamefont{L.}~\bibnamefont{Salem}}\ and\ \bibinfo
  {author} {\bibfnamefont{C.}~\bibnamefont{Rowland}},\ }%
  \bibfield{journal}{%
  \bibinfo {journal} {Angew. Chem. Int. Ed.}\ }%
  \textbf{\bibinfo {volume} {11}},\ \bibinfo {pages} {92} (\bibinfo {year}
  {1972})%
  \bibAnnoteFile{NoStop}{Salem1972}%
\bibitem{Comstock2005}%
  \BibitemOpen
  \bibfield{author}{%
  \bibinfo {author} {\bibfnamefont{M.}~\bibnamefont{Comstock}}, \bibinfo
  {author} {\bibfnamefont{J.}~\bibnamefont{Cho}}, \bibinfo {author}
  {\bibfnamefont{A.}~\bibnamefont{Kirakosian}},\ and\ \bibinfo {author}
  {\bibfnamefont{M.}~\bibnamefont{Crommie}},\ }%
  \bibfield{journal}{%
  \bibinfo {journal} {Phys. Rev. B}\ }%
  \textbf{\bibinfo {volume} {72}},\ \bibinfo {pages} {153414} (\bibinfo {year}
  {2005})%
  \bibAnnoteFile{NoStop}{Comstock2005}%
\bibitem{Vanderbilt1990}%
  \BibitemOpen
  \bibfield{author}{%
  \bibinfo {author} {\bibfnamefont{D.}~\bibnamefont{Vanderbilt}},\ }%
  \bibfield{journal}{%
  \bibinfo {journal} {Phys. Rev. B}\ }%
  \textbf{\bibinfo {volume} {41}},\ \bibinfo {pages} {7892} (\bibinfo {year}
  {1990})%
  \bibAnnoteFile{NoStop}{Vanderbilt1990}%
\bibitem{Monkhorst76}%
  \BibitemOpen
  \bibfield{author}{%
  \bibinfo {author} {\bibfnamefont{H.~J.}\ \bibnamefont{Monkhorst}}\ and\
  \bibinfo {author} {\bibfnamefont{J.~D.}\ \bibnamefont{Pack}},\ }%
  \bibfield{journal}{%
  \bibinfo {journal} {Phys. Rev. B}\ }%
  \textbf{\bibinfo {volume} {13}},\ \bibinfo {pages} {5188} (\bibinfo {year}
  {1976})%
  \bibAnnoteFile{NoStop}{Monkhorst76}%
\bibitem{Mercurio2010}%
  \BibitemOpen
  \bibfield{author}{%
  \bibinfo {author} {\bibfnamefont{G.}~\bibnamefont{Mercurio}}, \bibinfo
  {author} {\bibfnamefont{E.}~\bibnamefont{McNellis}}, \bibinfo {author}
  {\bibfnamefont{I.}~\bibnamefont{Martin}}, \bibinfo {author}
  {\bibfnamefont{S.}~\bibnamefont{Hagen}}, \bibinfo {author}
  {\bibfnamefont{F.}~\bibnamefont{Leyssner}}, \bibinfo {author}
  {\bibfnamefont{S.}~\bibnamefont{Soubatch}}, \bibinfo {author}
  {\bibfnamefont{J.}~\bibnamefont{Meyer}}, \bibinfo {author}
  {\bibfnamefont{M.}~\bibnamefont{Wolf}}, \bibinfo {author}
  {\bibfnamefont{P.}~\bibnamefont{Tegeder}}, \bibinfo {author}
  {\bibfnamefont{F.}~\bibnamefont{Tautz}},\ and\ \bibinfo {author}
  {\bibfnamefont{R.}~\bibnamefont{K.}},\ }%
  \bibfield{journal}{%
  \bibinfo {journal} {Phys. Rev. Lett.}\ }%
  \textbf{\bibinfo {volume} {104}},\ \bibinfo {pages} {36102} (\bibinfo {year}
  {2010})%
  \bibAnnoteFile{NoStop}{Mercurio2010}%
\bibitem{McNellis2010}%
  \BibitemOpen
  \bibfield{author}{%
  \bibinfo {author} {\bibfnamefont{E.~R.}\ \bibnamefont{McNellis}}, \bibinfo
  {author} {\bibfnamefont{C.}~\bibnamefont{Bronner}}, \bibinfo {author}
  {\bibfnamefont{J.}~\bibnamefont{Meyer}}, \bibinfo {author}
  {\bibfnamefont{M.}~\bibnamefont{Weinelt}}, \bibinfo {author}
  {\bibfnamefont{P.}~\bibnamefont{Tegeder}},\ and\ \bibinfo {author}
  {\bibfnamefont{K.}~\bibnamefont{Reuter}},\ }%
  \bibfield{journal}{%
  \bibinfo {journal} {Phys. Chem. Chem. Phys.: PCCP}\ }%
  \textbf{\bibinfo {volume} {12}},\ \bibinfo {pages} {6404} (\bibinfo {year}
  {2010})%
  \bibAnnoteFile{NoStop}{McNellis2010}%
\bibitem{Ruiz2012}%
  \BibitemOpen
  \bibfield{author}{%
  \bibinfo {author} {\bibfnamefont{V.}~\bibnamefont{Ruiz}}, \bibinfo {author}
  {\bibfnamefont{W.}~\bibnamefont{Liu}}, \bibinfo {author}
  {\bibfnamefont{E.}~\bibnamefont{Zojer}}, \bibinfo {author}
  {\bibfnamefont{M.}~\bibnamefont{Scheffler}},\ and\ \bibinfo {author}
  {\bibfnamefont{A.}~\bibnamefont{Tkatchenko}},\ }%
  \bibfield{journal}{%
  \bibinfo {journal} {Phys. Rev. Lett.}\ }%
  \textbf{\bibinfo {volume} {108}},\ \bibinfo {pages} {146103} (\bibinfo {year}
  {2012})%
  \bibAnnoteFile{NoStop}{Ruiz2012}%
\end{thebibliography}

\end{document}